\documentclass{iopart}

\newcommand{\be}{\begin{equation}}
\newcommand{\ee}{\end{equation}}

\begin{document}
\title[Fermion correlation functions in the massive Thirring
model]{Exact asymptotic behaviour of fermion correlation functions
in the massive Thirring model}

\author{Leonardo Mondaini and E C Marino}

\address{ Instituto de F\'{\i}sica, Universidade Federal do Rio de Janeiro,
Cx. Postal 68528, Rio de Janeiro RJ 21941-972, Brazil}

\eads{\mailto {mondaini@if.ufrj.br}, \mailto{marino@if.ufrj.br}}

\begin{abstract}
We obtain an exact asymptotic expression for the two-point fermion
correlation functions in the massive Thirring model (MTM) and show
that, for $\beta^2=8\pi$, they reproduce the exactly known
corresponding functions of the massless theory, explicitly
confirming the irrelevance of the mass term at this point. This
result is obtained by using the Coulomb gas representation of the
fermionic MTM correlators in the bipolar coordinate system.
\end{abstract}

\submitto{\JPA}

\section{Introduction}

In a recent paper \cite{MM1} we have presented the first exact
evaluation of the Kosterlitz-Thouless (KT) critical exponent
appearing in the asymptotic large distance behaviour of the
two-point spin correlation function of the XY-model (which is the
name given for the system consisting of planar spins interacting
through an exchange coupling in a lattice). This has been done by
using its connection to the sine-Gordon (SG) theory
\cite{KT,tsvelik} and the two-dimensional (2D) neutral Coulomb gas
(CG) \cite{sgcg} expressed in bipolar coordinates, which allow us to
obtain a convenient representation for the relevant correlator.

In this work we employ the same methodology established in
\cite{MM1} to compute the two-point fermion correlation functions of
the MTM, which, as is well-known, is also connected to the SG theory
\cite{col}. The MTM and the associated SG theory, indeed, are some of the best
studied quantum field theories. Numerous nontrivial exact results
have been obtained for this fascinating system. Among them, we may
list: the demonstration of the identity between the vacuum
functionals of the MTM and SG theory \cite{col}; the identification
of the fermionic MTM field as the soliton operator of the SG theory
\cite{col,mand}; the explicit obtainment of an expression for this
field operator in terms of the SG field (bosonization) \cite{mand};
the exact S-matrix and spectrum of bound-states \cite{zz}; recent
investigations on new aspects of the relationship between the 2D
Thirring model and the SG theory \cite{faber,faber2,juricic}; the
determination of the free energy and the specific heat of the system
by means of the thermodynamic Bethe ansatz \cite{dv}; and, finally,
the derivation of exact form factors for the soliton operators and
other fields \cite{sr1,sr2,babu1,babu2,babu3} and its consequent use
for the computation of density correlation functions \cite{sr6}.
Several exact results concerning the equilibrium statistical
mechanics of the 2D classical CG have also been obtained \cite{sr7}.
Among these, we mention the full thermodynamics for $0 \leq \beta^2
< 4\pi$, in the case of point particles \cite{sr4} and for $4\pi
\leq \beta^2 < 6\pi$, in the case of extensive ones \cite{sr5}.
Charge and particle correlators have been obtained in the low
temperature ($\beta^2 > 8\pi$) phase \cite{sr3}.

Renormalization group analysis of the SG/CG system has also produced
numerous interesting results. It has been shown, in particular, that
the mass term of the MTM or, equivalently, the cosine interaction of
the SG theory, becomes irrelevant for $\beta^2 \geq 8\pi$
\cite{amit,zj}. The continuous phase transition of Kosterlitz and
Thouless \cite{KosT} was identified in the associated XY-model of
spins at the temperature corresponding to this value of the SG
coupling $\beta$ and the associated critical exponent was evaluated
by using scaling arguments and the irrelevance of the corresponding
interaction \cite{KT,amit,jose,giam}.

Despite this huge mass of important results, however, the fermion
field correlation functions of the MTM are not known exactly, except
for the special point $\beta^2=4\pi$ \cite{le}, where the MTM
becomes a free massive theory. It is the purpose of this work to
obtain an exact large distance asymptotic expression for the fermion
correlators of the MTM and to show that for $\beta^2=8\pi$ (KT
critical point) this asymptotic behaviour reproduces the exactly
known fermion correlation functions of the massless Thirring model
\cite{klaiber}, thus confirming explicitly the irrelevance of the
mass term of the MTM at this point. In order to do that, we make use
of the CG representation of the SG system and the bosonized form of
the MTM fields, which in the CG framework become associated with
external charges and strings of electric dipoles interacting with
the charges of the gas. The exact large distance asymptotic form of
the correlators is then obtained by the use of a special coordinate
system, namely, the bipolar coordinates.

\section{The MTM and the SG/CG system}

In this section, we are going to review some basic features of the
connection of the MTM with the SG theory and with the 2D neutral CG.
We then finish by presenting a representation of the fermionic MTM
correlators in the framework of the classical CG, which will be our
starting point for their evaluation at $\beta^2=8\pi$.

The MTM is described by the lagrangian
\begin{equation}
{\cal L} = \rmi \bar \psi\not\!\partial\psi - M_0 \bar \psi \psi -
\frac{g}{2}(\bar \psi \gamma_\mu \psi)(\bar \psi \gamma^\mu \psi) ,
\label{1}
\end{equation}
where $\psi$ is a two-component Dirac fermion field in 1+1D. It is
well known that it can be mapped into the SG theory of an scalar
field \cite{col} whose dynamics is determined by
\begin{equation}
{\cal L} = \frac{1}{2}\partial_\mu\phi\
\partial^\mu\phi +2\alpha_0 \cos\beta\phi ,
\label{2}
\end{equation}
where the couplings in the two models are related as:
\begin{equation}
g = \pi \left (\frac{4\pi}{\beta^2} - 1 \right) , \ \ \ \ \ \ \ \ \
\ \ \ M_0 \bar \psi \psi = - 2\alpha_0 \cos\beta\phi . \label{2a}
\end{equation}
Under this mapping, the two components of the fermion field may be
expressed in terms of the SG field as
\begin{equation}
\psi_{1}(\vec x) =  \sigma(\vec x)\mu(\vec x) , \ \ \ \ \ \ \ \ \ \
\ \ \psi_{2}(\vec x) = \sigma^\dagger(\vec x)\mu(\vec x),
\label{3}
\end{equation}
where $\sigma(\vec x)$ and $\mu(\vec x)$ are, respectively, order
and disorder fields, satisfying a dual algebra, which can be
introduced in the SG theory \cite{ms}. These are given by
\begin{equation}
\sigma(x,\tau) = \exp \left\{ \rmi\  \frac{\beta}{2} \ \phi (x,\tau)
\right\},
\label{4}
\end{equation}
\begin{equation}
\mu(x,\tau) = \exp \left\{ \rmi\ \frac{2\pi}{\beta}
\int_{-\infty}^{x} \rmd z\, \dot{\phi} (z,\tau) \right\}.
\label{5}
\end{equation}
\Eref{3} coincides with the bosonized expression for the fermion
field, first obtained in \cite{mand}.

In what follows, we are going to perform an expansion in $\alpha_0$.
In order to control the infrared (IR) divergences inherent to the
expansion around a massless theory in 2D, we follow \cite{col} and
modify (\ref{2}) by adding a regulator mass term
\begin{equation}
{\cal L}_{reg} = \frac{1}{2}\mu^2_0 \phi^2 \label{2a}
\end{equation}
and multiplying the interaction term $(2\alpha_0 \cos\beta\phi)$ by
a function $f(\vec z)$ of compact support. At the end, of course, we
must take the limits $\mu_0\rightarrow 0$ and $f(\vec z)\rightarrow
1$.

 The euclidean vacuum functional of the SG theory may be written
as the grand-partition function of a classical neutral 2D CG, namely
\cite{sgcg,amit}
\begin{eqnarray}
\fl\mathcal{Z}=\lim_{\varepsilon\rightarrow 0}\lim_{f(z)\rightarrow
1}\lim_{\mu_0\rightarrow 0}
\sum_{n=0}^{\infty}\frac{\alpha^{2n}}{(n!)^2}\int
\prod_{i=1}^{2n}(\rmd^2z_i\, f(\vec z_i)) \nonumber
\\ \times\exp\left\{\frac{\beta^2}{8\pi}\sum_{i\neq
j=1}^{2n}\lambda_i\lambda_j\ln \left[\mu_0^2\left(|\vec z_i-\vec
z_j|^2+|\varepsilon|^2\right)\right]\right\}, \label{6}
\end{eqnarray}
where $\lambda_i = 1$ for $1 \leq i \leq n$ and $\lambda_i = - 1$
for $n+1 \leq i \leq 2n$  and $\varepsilon$ is a ultraviolet (UV)
regulator, introduced in the 2D Coulomb potential, which is needed
in the case of point particles or, equivalently, of a local field
theory. The renormalized coupling $\alpha$ is related to the one in
(\ref{2}) by
\begin{equation}
\alpha = \alpha_0
\left(\mu_0^2|\varepsilon|^2\right)^{\frac{\beta^2}{8\pi}} .
\label{7}
\end{equation}

We must emphasize that, in order to obtain (\ref{6}), use was made
of the UV and IR regulated Green's function of the free scalar
theory, namely \cite{amit}
\begin{equation}
\fl G(\vec r;\mu_0) = \frac{1}{2\pi} {\rm {K}}_0 \left[\mu_0
\left(|\vec r|^2 +|
\varepsilon|^2\right)^{\frac{1}{2}}\right]\stackrel{\mu_0 |\vec
r|\ll 1 }{\sim} -\frac{1}{4\pi} \ln \left[\mu^2_0 \left(|\vec r|^2
+| \varepsilon|^2\right)\right],\label{7a}
\end{equation}
where ${\rm {K}}_0$ is a Bessel function.

Notice that, due to neutrality, the explicit dependence on $\mu_0$
disappears from the summand in (\ref{6}).

In the CG language, the couplings $\alpha$ and $\beta$ are related,
respectively, to the CG fugacity and temperature as $\alpha
=\mu_{CG}$ and $\beta^2=\frac{2\pi}{k_BT_{CG}}$. At the
Kosterlitz-Thouless point $T_{KT}$, corresponding to $\beta^2=8\pi$,
the system undergoes a phase transition from a metallic (fluid)
phase composed of charged particles, for $\beta^2<8\pi$, to an
insulating (dielectric) phase, composed of neutral dipoles, for
$\beta^2>8\pi$. In the region $0< \beta^2 < 4\pi$ the singularities
occurring in (\ref{6}) are all integrable and no UV regularization
is needed for the Green's function. The CG of point charges is
thermodynamically stable. For $4\pi \leq \beta^2 < 8\pi$, however,
the singularities are no longer integrable and the system becomes
unstable. Use of the UV regularized Greens's function introduced
above becomes therefore necessary, in order to prevent thermodynamic
collapse.

Using the CG description we can write the four components of the
two-point fermion correlation function as
\begin{eqnarray}
\fl \langle\psi_{1(2)}(\vec x)\psi_{1(2)}^\dagger(\vec y)\rangle =
\lim_{\varepsilon\rightarrow 0}\lim_{f(z)\rightarrow
1}\lim_{\mu_0\rightarrow 0} \frac{+(-)\rmi \ \exp
\left[+(-)\rmi\arg(\vec x-\vec y)\right]}{\mathcal{Z}} \left[
\frac{|\varepsilon|} {|\vec x-\vec
y|}\right]^{\left(\frac{2\pi}{\beta^2}+\frac{\beta^2}{8\pi}\right)}
\nonumber \\ \times
\sum_{n=0}^{\infty}\frac{\alpha^{2n}}{(n!)^2}\int
\prod_{i=1}^{2n}(\rmd^2z_i\, f(\vec z_i)) \nonumber\\  \times\exp
\left \{\frac{\beta^2}{8\pi}\sum_{i\neq
j=1}^{2n}\lambda_i\lambda_j\ln \left[\mu_0^2\left(|\vec z_i-\vec
z_j|^2+|\varepsilon|^2\right)\right]\nonumber \right . \\
\left .+(-) \frac{\beta^2}{8\pi}\sum_{i=1}^{2n}\lambda_i\ln
\frac{\left[|\vec z_i-x|^2 + |\varepsilon|^2\right]}{\left[|\vec z_i-y|^2 + |\varepsilon|^2\right]}\nonumber \right . \\
\left .+\rmi\sum_{i=1}^{2n} \lambda_i\lbrack \arg(\vec z_i-\vec
y)-\arg(\vec z_i-\vec x)\rbrack \right \}
\label{8}
\end{eqnarray}
and
\begin{eqnarray}
\fl \langle\psi_{1(2)}(\vec x)\psi_{2(1)}^\dagger(\vec y)\rangle =
\lim_{\varepsilon\rightarrow 0}\lim_{f(z)\rightarrow
1}\lim_{\mu_0\rightarrow 0}
\frac{-(+)\rmi}{\mathcal{Z}}\left[\mu_0|\varepsilon|\right]^{\left(\frac{2\pi}{\beta^2}
+\frac{\beta^2}{8\pi}\right)} \left[\mu_0|\vec x-\vec
y|\right]^{-\left(\frac{2\pi}{\beta^2}-\frac{\beta^2}{8\pi}\right)}
\nonumber\\ \times
\sum_{n=0}^{\infty}\frac{\alpha^{(2n+1)}}{n!(n+1)!}\int
\prod_{i=1}^{2n+1}(\rmd^2z_i\, f(\vec z_i)) \nonumber\\ \times\exp
\left \{\frac{\beta^2}{8\pi}\sum_{i\neq
j=1}^{2n+1}\lambda_i\lambda_j\ln \left[\mu_0^2\left(|\vec z_i-\vec
z_j|^2+|\varepsilon|^2\right)\right]\nonumber \right . \\
\left .+(-)\frac{\beta^2}{8\pi}\sum_{i=1}^{2n+1}\lambda_i\ln \left
\{ \left[\mu_0^2\left(|\vec z_i-\vec
x|^2+|\varepsilon|^2\right)\right] \left[\mu_0^2\left(|\vec z_i-\vec
y|^2+|\varepsilon|^2\right)\right]\right \}\nonumber \right . \\
\left .+\rmi\sum_{i=1}^{2n+1} \lambda_i\lbrack \arg(\vec z_i-\vec
y)-\arg(\vec z_i-\vec x)\rbrack \right \} . \label{9}
\end{eqnarray}

In (\ref{8}) we have $\lambda_i = +1$ for $1 \leq i \leq n$ and
$\lambda_i = - 1$ for $n+1 \leq i \leq 2n$. In (\ref{9}), on the
other hand, we have $\lambda_i = +1$ for $1 \leq i \leq n$ and
$\lambda_i = - 1$ for $n+1 \leq i \leq 2n+1$ for
$\langle\psi_{1}\psi_{2}^\dagger\rangle$. Conversely, for
$\langle\psi_{2}\psi_{1}^\dagger\rangle$, we have $\lambda_i = -1$
for $1 \leq i \leq n$ and $\lambda_i = + 1$ for $n+1 \leq i \leq
2n+1$.

Notice that the order fields introduce additional external charges
of half magnitude in the gas. The disorder fields, on the other
hand, introduce strings of electric dipoles connecting $\vec x$ and
$\vec y$ and whose interaction potential with a charge at $\vec z$
is proportional to $\left[\arg(\vec z-\vec y)-\arg(\vec z-\vec
x)\right]$ \cite{ms}. As a consequence, in the case of the diagonal
components of the fermionic correlators, we have two external
charges with half of the magnitude of the gas charges and {\it
opposite} signs, located at $\vec x$ and $\vec y$. The CG,
therefore, remains neutral. In the case of the off-diagonal
components, however, the fermion fields introduce two external
charges, also having half magnitude and with the {\it same} sign at
$\vec x$ and $\vec y$. In order to achieve global neutrality,
therefore, the CG must be no longer neutral, having an extra
positive charge in the case of
$\langle\psi_{2}\psi_{1}^\dagger\rangle$ and an extra negative
charge in the case of $\langle\psi_{1}\psi_{2}^\dagger\rangle$.
Global neutrality is a necessary condition for the existence of the
$\mu_0 \rightarrow 0$ limit, since in this case the $\mu_0$-factors
are completely canceled in (\ref{8}) and (\ref{9}).

\section{Bipolar coordinates and the fermion correlators}

In this section we make use of bipolar coordinates in order to obtain a
representation for the fermion correlators that will prove to be extremely
useful. It allows, in particular, the obtainment of a series in $|\vec x - \vec y|$
for these correlators, valid for $\vec x \neq \vec y$, out of which we can derive
an exact asymptotic expression at $\beta^2 = 8\pi$.

Given the
position vector $\vec r$ in the plane and two points (poles) at
$\vec x$ and $\vec y$, the bipolar coordinates ($\xi,\eta$) are defined as
\cite{arf}
\begin{equation}
\xi = \arg (\vec r  - \vec y) - \arg (\vec r  - \vec x) , \ \ \ \ \
\ \ \ \ \ \ \ \eta = \ln \frac{|\vec r  - \vec x|}{|\vec r  - \vec
y|} , \label{10}
\end{equation}
with $0 \leq \xi \leq 2\pi$ and $-\infty < \eta < \infty$. In terms
of these coordinates, the position vector is given by
\begin{equation}
\vec r = \frac{|\vec x  - \vec y| }{2 [\cosh \eta -\cos \xi] }\left
(\sinh \eta \ , \
 \sin \xi \right )
\label{11}
\end{equation}
and the volume element reads
\begin{equation}
\rmd^2z = \frac{|\vec x - \vec y|^2}{4 [\cosh \eta -\cos \xi]^2 }
\rmd\xi \rmd\eta. \label{12}
\end{equation}

Rewriting expressions (\ref{8}) and (\ref{9}), for $\vec x \neq \vec y$, in terms of
bipolar coordinates, we get
\begin{eqnarray}
\fl \langle\psi_{1(2)}(\vec x)\psi_{1(2)}^\dagger(\vec y)\rangle =
\lim_{\varepsilon\rightarrow 0}\lim_{f(z)\rightarrow
1}\lim_{\mu_0\rightarrow 0} \frac{+(-)\rmi \ \exp
\left[+(-)\rmi\arg(\vec x-\vec y)\right]}{\mathcal{Z}} \left[
\frac{|\varepsilon|} {|\vec x-\vec
y|}\right]^{\left(\frac{2\pi}{\beta^2}+\frac{\beta^2}{8\pi}\right)}
\nonumber \\ \times \sum_{n=0}^{\infty}\frac{\alpha^{2n}}{(n!)^2}
\int_{0,V(\varepsilon)}^{2\pi}
\int_{-\infty,V(\varepsilon)}^{+\infty} \prod_{i=1}^{2n} (\rmd\xi_i
\rmd\eta_i\, f(\xi_i,\eta_i)) \nonumber\\ \times\frac{|\vec x  -
\vec y|^{4n}}{4 [\cosh \eta_i -\cos \xi_i]^2 } \exp \left
\{\frac{\beta^2}{8\pi}\sum_{i\neq j=1}^{2n}\lambda_i\lambda_j
\nonumber\right .\\
\left .\times\ln \left \{ \left[\mu_0|\vec x  - \vec y|\right]^{2}
\left[ \left ( \frac{\sinh \eta_i}{2 [\cosh \eta_i -\cos \xi_i]}-
\frac{ \sinh \eta_j}{2 [\cosh
\eta_j -\cos \xi_j]} \right )^2\nonumber\right .\right .\right .\\
\left . \left .\left . + \left ( \frac{\sin \xi_i}{2 [\cosh \eta_i
-\cos \xi_i]}- \frac{ \sin \xi_j}{2 [\cosh \eta_j -\cos \xi_j]}
\right )^2 \right ] \right \}
\nonumber\right .\\
\left . +(-)\frac{\beta^2}{4\pi}\sum_{i=1}^{2n} \lambda_i
\eta_i+\rmi\sum_{i=1}^{2n} \lambda_i \xi_i \right \} ,
\label{13}
\end{eqnarray}
and
\begin{eqnarray}
\fl \langle\psi_{1(2)}(\vec x)\psi_{2(1)}^\dagger(\vec y)\rangle =
\lim_{\varepsilon\rightarrow 0}\lim_{f(z)\rightarrow
1}\lim_{\mu_0\rightarrow 0}
\frac{-(+)\rmi}{\mathcal{Z}}\left[\mu_0|\varepsilon|\right]^{\left(\frac{2\pi}{\beta^2}
+\frac{\beta^2}{8\pi}\right)} \left[\mu_0|\vec x-\vec
y|\right]^{-\left(\frac{2\pi}{\beta^2}-\frac{\beta^2}{8\pi}\right)}
\nonumber\\ \times
\sum_{n=0}^{\infty}\frac{\alpha^{(2n+1)}}{n!(n+1)!}
\int_{0,V(\varepsilon)}^{2\pi}
\int_{-\infty,V(\varepsilon)}^{+\infty} \prod_{i=1}^{2n+1}
(\rmd\xi_i \rmd\eta_i\, f(\xi_i,\eta_i)) \nonumber\\
\times\frac{|\vec x  - \vec y|^{(4n+2)}}{4 [\cosh \eta_i -\cos
\xi_i]^2 } \exp \left \{\frac{\beta^2}{8\pi}\sum_{i\neq
j=1}^{2n+1}\lambda_i\lambda_j
\nonumber\right .\\
\left .\times\ln \left \{ \left[\mu_0|\vec x  - \vec y|\right]^{2}
\left[ \left ( \frac{\sinh \eta_i}{2 [\cosh \eta_i -\cos \xi_i]}-
\frac{ \sinh \eta_j}{2 [\cosh
\eta_j -\cos \xi_j]} \right )^2\nonumber\right .\right .\right .\\
\left . \left .\left . + \left ( \frac{\sin \xi_i}{2 [\cosh \eta_i
-\cos \xi_i]}- \frac{ \sin \xi_j}{2 [\cosh \eta_j -\cos \xi_j]}
\right )^2 \right ] \right \}
\nonumber\right .\\
\left . +(-)\frac{\beta^2}{8\pi}\sum_{i=1}^{2n+1} \lambda_i \ln
\left \{ \left[\mu_0|\vec x  - \vec y|\right]^{4} \left[ \left (
\frac{\sinh \eta_i}{2 [\cosh
\eta_i -\cos \xi_i]}+ \frac{1}{2} \right )^2\nonumber\right .\right .\right .\\
\left . \left .\left . + \left ( \frac{\sin \xi_i}{2 [\cosh \eta_i
-\cos \xi_i]} \right )^2 \right ] \left[ \left ( \frac{\sinh
\eta_i}{2 [\cosh
\eta_i -\cos \xi_i]}- \frac{1}{2} \right )^2\nonumber\right .\right .\right .\\
\left . \left .\left . + \left ( \frac{\sin \xi_i}{2 [\cosh \eta_i
-\cos \xi_i]} \right )^2 \right ]\right \}+\rmi\sum_{i=1}^{2n+1}
\lambda_i \xi_i \right \} .
\label{14}
\end{eqnarray}

Notice that in the two previous expressions, we modified the UV
regulating method, by redefining the integration region as
$V(\varepsilon)$, in such a way that the integrations must respect
the condition $|\vec z_i-\vec z_j| > \varepsilon$. In terms of the
$\xi_i$,$\eta_i$ integrals, this implies the following restriction
for the expressions between round brackets in (\ref{13}) and
(\ref{14}), which we call, respectively $\alpha_{ij} $ and
$\beta_{ij}$:
\begin{equation}
\left [  \alpha^2_{ij} + \beta^2_{ij}\right ] >
\frac{|\varepsilon|^2}{|\vec x - \vec y|^{2}}. \label{15}
\end{equation}

It is easy to see that the $|\vec x - \vec y|$-factors decouple from
the integrals in (\ref{13}) and (\ref{14}). We can also see that,
for finite $|\vec x - \vec y|$, the $\mu_0$-factors completely
cancel out from the fermion correlators, as we observed at the end
of the previous section.

A simple combinatoric analysis, considering the neutrality of the
system, shows that for the diagonal components of the correlation
functions, the $|\vec x  - \vec y|^{\frac{\beta^2}{2\pi}}$-factors
appear $n(n-1)$ times in the numerator and $n^2$ times in the
denominator. Adding the $4n$ contribution coming from the scale
factors of the volume elements, we obtain
\begin{eqnarray}
\fl \langle\psi_{1(2)}(\vec x)\psi_{1(2)}^\dagger(\vec y)\rangle =
\lim_{\varepsilon\rightarrow 0}\lim_{f(z)\rightarrow
1}\lim_{\mu_0\rightarrow 0} \frac{+(-)\rmi \ \exp
\left[+(-)\rmi\arg(\vec x-\vec y)\right]}{\mathcal{Z}} \left[
\frac{|\varepsilon|} {|\vec x-\vec
y|}\right]^{\left(\frac{2\pi}{\beta^2}+\frac{\beta^2}{8\pi}\right)}
\nonumber \\
\times \sum_{n=0}^{\infty}C_n^{+(-)} \left(|\vec x-\vec y|\right)\
|\vec x  - \vec y|^{\left(2 -\frac{\beta^2}{4\pi}\right)2n},
\label{16}
\end{eqnarray}
where the $C_n^{+(-)}$ coefficients are given by each term of the
summand in (\ref{13}) after the removal of the $|\vec x - \vec
y|$-factors.

Conversely, for the off-diagonal components, a similar combinatoric
analysis indicates that the $|\vec x  - \vec
y|^{\frac{\beta^2}{2\pi}}$-factors appear $n(n+1)$ times in the
numerator and $(n+1)^2$ times in the denominator. Considering the
$4n+2$ scale factors of the volume elements in (\ref{14}) we get
\begin{eqnarray}
\fl \langle\psi_{1(2)}(\vec x)\psi_{2(1)}^\dagger(\vec y)\rangle =
\lim_{\varepsilon\rightarrow 0}\lim_{f(z)\rightarrow
1}\lim_{\mu_0\rightarrow 0}
\frac{-(+)\rmi}{\mathcal{Z}}\left[\mu_0|\varepsilon|\right]^{\left(\frac{2\pi}{\beta^2}
+\frac{\beta^2}{8\pi}\right)} \left[\mu_0|\vec x-\vec
y|\right]^{-\left(\frac{2\pi}{\beta^2}-\frac{\beta^2}{8\pi}\right)}
\nonumber \\
\times \sum_{n=0}^{\infty}F_n^{+(-)} \left(|\vec x-\vec y|\right)\
|\vec x  - \vec y|^{\left[\left(2 -\frac{\beta^2}{4\pi}\right)2n -
\frac{\beta^2}{2\pi} + 2\right]}, \label{17}
\end{eqnarray}
where the $F_n^{+(-)}$ coefficients are given by each term of the
summand in (\ref{14}) after the removal of the $|\vec x - \vec
y|$-factors.

Notice that the coefficients $C_n^{+(-)}$ and $F_n^{+(-)}$ in
(\ref{16}) and (\ref{17}) depend on $|\vec x  - \vec y|$ through the
restriction on the integration region given by (\ref{15}).

In the limit $\alpha \rightarrow 0$, our expressions for the fermion
correlation functions of the MTM reproduce the exact solution for
the euclidean correlators of the massless Thirring model. In the
limit $\beta \rightarrow 0$ the mass operator becomes trivial and,
again, we must take $\alpha \rightarrow 0$, thereby recovering the
exact correlation functions of the massless Thirring model.

It is also interesting to note that our expressions for the
two-point fermion correlators of the MTM, (\ref{16}) and (\ref{17})
reproduce, in the case $\beta^2 = 4\pi$ (free-fermion point) the
free massive fermion correlation function
\begin{equation}
\fl \langle\psi(\vec x)\psi^\dagger(\vec y)\rangle_0 = M_0 \left(
\begin{array}{cc}
\zeta \ {\rm {K}}_1 (M_0 |\vec x-\vec y|) & {\rm {K}}_0 (M_0 |\vec x-\vec y|)  \\
\, & \, \\
{\rm {K}}_0 (M_0 |\vec x-\vec y|) & \zeta^\ast \ {\rm {K}}_1 (M_0
|\vec x-\vec y|)
\end{array} \right),
\end{equation}
where $\zeta = \rmi \rme^{\rmi \arg(\vec x-\vec y)}$ and $M_0$ is
the free fermion mass. Indeed, for $\beta^2 = 4\pi$ the series
appearing in (\ref{16}) and (\ref{17}), respectively, are precisely
the ones that occur in the definition of the Bessel functions ${\rm
{K}}_1$ and ${\rm {K}}_0$ \cite{grad}. Equating the coefficients we
obtain the following expressions for $C_n^{+(-)}$ and $F_n^{+(-)}$
($C_0^{+(-)}\equiv 1$ for all values of $\beta$):
\begin{equation}
\fl C_{n+1}^{+(-)}(|\vec x-\vec y|) =
\frac{{M_0}^{2n+2}}{2^{2n+1}n!(n+1)!}\left[\ln\left( \frac{M_0|\vec
x-\vec y|}{2} \right) -\frac{1}{2}\psi(n+1) -\frac{1}{2} \psi(n+2)
\right ]
 \label{15a}
\end{equation}
and
\begin{equation}
 F_n^{+(-)}(|\vec x-\vec y|) = \frac{+(-)\rmi \, {M_0}^{2n}}{2^{2n}(n!)^2}\left[\psi(n+1)-\ln\left(
\frac{M_0|\vec x-\vec y|}{2} \right) \right ]
 \label{15b},
\end{equation}
where $\psi(x)$ is the Euler function. In order to obtain
(\ref{15a}) and (\ref{15b}), we used the fact that (\ref{16}) and
(\ref{17}) are independent of $\mu_0$. We then replaced $\mu_0$ for
the physical mass $M_0$ and eliminated the $|\varepsilon|$ and  $\mathcal{Z}$-factors by
renormalizing the fermion fields.

From the exact expression (\ref{16}), we clearly see a definite
change in the large distance behaviour of the diagonal correlation
functions at $\beta^2 = 8\pi$. This indicates that for $\beta^2 >
8\pi$, the asymptotic large distance behaviour is determined by the
corresponding massless correlator. We are going to see, in the next
section, that for these values of the coupling constant $\beta$,
also the off-diagonal components of the correlator at large
distance, correspond to the respective massless correlators, namely,
they vanish identically.

\section{Asymptotic behaviour of fermion correlators}

\subsection{Diagonal components}

Let us study in this subsection the large distance limit of the diagonal, chirality conserving,
components of the fermion correlation function at $\beta^2 = 8\pi$. From (\ref{16}), we can write
\begin{eqnarray}
\fl \langle\psi_{1(2)}(\vec x)\psi_{1(2)}^\dagger(\vec y)\rangle =
\lim_{\varepsilon\rightarrow 0}\lim_{f(z)\rightarrow
1}\lim_{\mu_0\rightarrow 0} +(-)\rmi \ \exp \left[+(-)\rmi\arg(\vec
x-\vec y)\right] \left[ \frac{|\varepsilon|} {|\vec x-\vec
y|}\right]^{\frac{5}{4}}\nonumber \\ \times\ K^{+(-)} \left(|\vec
x-\vec y|\right) , \label{18}
\end{eqnarray}
where
\begin{equation}
K^{+(-)} \left(|\vec x-\vec y|\right) = \mathcal{Z}^{-1}
\sum_{n=0}^{\infty} C_n^{+(-)} \left(|\vec x-\vec y|\right)
\label{19}
\end{equation}
evaluated at $\beta^2 = 8\pi$.

Going back to the original coordinate system and defining the symbols
\begin{equation}
[ x_i ,y_j] \equiv \mu_0^4\left[|\vec x_i  - \vec
y_j|^2+|\varepsilon|^2\right]^2 , \label{20}
\end{equation}
we may express $K^{+(-)} \left(|\vec
x-\vec y|\right)$ in the form
\begin{eqnarray}
\fl K^{+(-)} \left(|\vec x-\vec y|\right) =  \mathcal{Z}^{-1}
\sum_{n=0}^{\infty}\frac{\alpha^{2n}}{(n!)^2}\int
\prod_{i=1}^{n}(\rmd^2x_i\, f(\vec x_i)) \prod_{i=1}^{n}(\rmd^2y_i\, f(\vec y_i)) \nonumber \\
\times \frac{\prod_{i<j}^{n} \frac{[x_i , x_j]}{|\vec x - \vec y|^4}
\prod_{i<j}^{n}  \frac{[ y_i , y_j]}{|\vec x - \vec y|^4}}{|\vec x -
\vec y|^{4n} \prod_{i,j}^{n} \frac{[ x_i , y_j]}{|\vec x - \vec
y|^4}} \nonumber \\ \times \left (\frac{\prod_{i}^{n} [x_i ,x]
\prod_{i}^{n}  [ y_i ,y]} {\prod_{i}^{n} [ x_i
,y] \prod_{i}^{n} [y_i ,x]}\right)^{+(-)\frac{1}{2}} \nonumber \\
\times \exp \left\{\rmi\sum_{i=1}^{n} \lbrack \arg(\vec
x_i-\vec y)-\arg(\vec x_i-\vec x)\rbrack \nonumber \right . \\
\left . - \rmi\sum_{i=1}^{n} \lbrack \arg(\vec y_i-\vec y)-\arg(\vec
y_i-\vec x)\rbrack\right \} , \label{21}
\end{eqnarray}
where we went back to the original UV regulating method and
associated the positive charges with $\vec x_i$ and the negative
ones with $\vec y_i$. Observe that the $|\vec x-\vec y|$-factors
completely cancel out in (\ref{21}) and, therefore, may be removed.

Let us now study the asymptotic large distance behaviour of
(\ref{18}). In order to do that we must rewrite the expression
between round brackets, as well as the phases, in (\ref{21}) in the
form that we would have obtained if we had used the fully regulated
form of the Green's function, given in (\ref{7a}), since in this
limit the last part of that expression is no longer valid. Thus, we
should write the expression between round brackets as
\begin{eqnarray}
\fl \exp\left\{-(+)2\left[\sum_{i=1}^{n}\left({\rm {K}}_0
\left[\mu_0\left(|\vec x_i-\vec
x|^2+|\varepsilon|^2\right)^{\frac{1}{2}}\right]- {\rm {K}}_0
\left[\mu_0\left(|\vec x_i-\vec
y|^2+|\varepsilon|^2\right)^{\frac{1}{2}}\right]\right)\nonumber\right .\right .\\
\left .\left .\fl -\sum_{i=1}^{n}\left( {\rm {K}}_0
\left[\mu_0\left(|\vec y_i-\vec
x|^2+|\varepsilon|^2\right)^{\frac{1}{2}}\right]- {\rm {K}}_0
\left[\mu_0\left(|\vec y_i-\vec
y|^2+|\varepsilon|^2\right)^{\frac{1}{2}}\right]\right)\right]\right\},
\label{21b}
\end{eqnarray}
whereas, for the phases, we get
\begin{eqnarray}
\fl \int_{\vec y}^{2\vec y-\vec x}\rmd\xi_{\mu}\,
\epsilon^{\mu\nu}\partial_{\nu}^{(\xi)}\left(\sum_{i=1}^{n}{\rm
{K}}_0 \left[\mu_0\left(|\vec \xi-\vec x_i+(\vec x-\vec
y)|^2+|\varepsilon|^2\right)^{\frac{1}{2}}\right]\right .\nonumber
\\ \left .-\sum_{i=1}^{n}{\rm {K}}_0 \left[\mu_0\left(|\vec \xi-\vec
y_i+(\vec x-\vec
y)|^2+|\varepsilon|^2\right)^{\frac{1}{2}}\right]\right) ,
\label{21c}
\end{eqnarray}
where we have performed the shift $\vec \xi \rightarrow \vec
\xi-(\vec x - \vec y)$ in the integration variable. Notice that the
former expression for the phases may be obtained from (\ref{21c}),
for $\mu_0 |\vec r| \ll 1 $, by using (\ref{7a}) and the
Cauchy-Riemann equation for the logarithm function \cite{ms}.

It is easy to see that, for $|\vec x-\vec y| \rightarrow \infty$, we
have
\begin{equation}
[ x_i ,x] \stackrel{|\vec x-\vec y| \rightarrow \infty}{\sim} [ x_i
,y] , \ \ \ \ \ \ \ \ \ \ \ \ [y_i ,x]\stackrel{|\vec x-\vec y|
\rightarrow \infty}{\sim} [ y_i ,y] \label{22}
\end{equation}
and therefore, the expression (\ref{21b}) tends to one. Using the
fact that ${\rm {K}}_0(x)\stackrel{x \rightarrow
\infty}{\longrightarrow} 0$ we may also see that the phases
(\ref{21c}) vanish in the large distance limit. Consequently,
considering that the remaining terms in the summand in (\ref{21})
are identical to those in $\mathcal{Z}$ we get
\begin{equation}
K^{+(-)} \left(|\vec x-\vec y|\right) \stackrel{|\vec x-\vec y|
\rightarrow \infty}{\longrightarrow} 1 . \label{23}
\end{equation}
The IR regulator, $\mu_0$, as well as the functions $f(\vec z)$, can
now be safely removed in (\ref{18}). Introducing the renormalized
fields
\begin{equation}
\psi_{1(2)}^{\rm{R}} = \psi_{1(2)}\ |\varepsilon|^{-\frac{5}{8}} ,
\label{24}
\end{equation}
also the UV regulator $\varepsilon$ may be removed in (\ref{18}) and
we, finally, obtain
\begin{equation}
\langle\psi_{1(2)}(\vec x)\psi_{1(2)}^\dagger(\vec
y)\rangle_{\rm{R}} \stackrel{|\vec x-\vec y| \rightarrow
\infty}{\sim} \frac{+(-)\rmi \ \exp \left[+(-)\rmi\arg(\vec x-\vec
y)\right]}{|\vec x-\vec y|^{\frac{5}{4}}}, \label{25}
\end{equation}
which are the diagonal
components of the euclidean correlator corresponding to Klaiber's
exact solution of the massless Thirring model \cite{klaiber}.

\subsection{Off-diagonal components}

We now consider the off-diagonal, chirality nonconserving, components
of the fermion correlation function at $\beta^2 = 8\pi$. From (\ref{17})
we may write
\begin{eqnarray}
\fl \langle\psi_{1(2)}(\vec x)\psi_{2(1)}^\dagger(\vec y)\rangle =
\lim_{\varepsilon\rightarrow 0}\lim_{f(z)\rightarrow
1}\lim_{\mu_0\rightarrow 0} -(+)\rmi
\left[\mu_0|\varepsilon|\right]^{\frac{5}{4}} \left[\mu_0|\vec
x-\vec y|\right]^{\frac{3}{4}} G^{+(-)} \left(|\vec x-\vec y|\right)
, \label{26}
\end{eqnarray}
where
\begin{equation}
G^{+(-)} \left(|\vec x-\vec y|\right) = \mathcal{Z}^{-1}\ |\vec x-\vec y|^{-2}
\sum_{n=0}^{\infty} F_n^{+(-)} \left(|\vec x-\vec y|\right)
\label{27}
\end{equation}
evaluated at $\beta^2 = 8\pi$.

Proceeding as before, we go back to the original coordinate system and express $G^+$
as
\begin{eqnarray}
\fl G^+ \left(|\vec x-\vec y|\right) =  \mathcal{Z}^{-1}\ |\vec
x-\vec y|^{-2}
\sum_{n=0}^{\infty}\frac{\alpha^{(2n+1)}}{n!(n+1)!}\int\prod_{i=1}^{n}(\rmd^2x_i\,
f(\vec x_i)) \prod_{i=1}^{n+1}(\rmd^2y_i\, f(\vec y_i)) \nonumber \\
\times \frac{\prod_{i<j}^{n} \frac{[x_i , x_j]}{|\vec x - \vec y|^4}
\prod_{i<j}^{n+1}  \frac{[ y_i , y_j]}{|\vec x - \vec y|^4}}{|\vec x
- \vec y|^{(4n+2)} \prod_{i}^{n}\prod_{j}^{n+1} \frac{[ x_i ,
y_j]}{|\vec x - \vec y|^4}} \nonumber
\\ \times \left( \frac{\prod_{i}^{n} \frac{[x_i , x]}{|\vec x - \vec y|^4} \prod_{i}^{n}
\frac{[x_i , y]}{|\vec x - \vec y|^4}}{ \prod_{i}^{n+1}\frac{[y_i ,
x]}{|\vec x - \vec y|^4} \prod_{i}^{n+1} \frac{[y_i , y]}{|\vec x -
\vec y|^4}}\right)^{\frac{1}{2}} \nonumber \\ \times \exp
\left\{\rmi\sum_{i=1}^{n} \lbrack \arg(\vec
x_i-\vec y)-\arg(\vec x_i-\vec x)\rbrack \nonumber \right . \\
\left . - \rmi\sum_{i=1}^{n+1} \lbrack \arg(\vec y_i-\vec
y)-\arg(\vec y_i-\vec x)\rbrack\right \} . \label{28}
\end{eqnarray}
$G^-$, accordingly, may be obtained from (\ref{28}) by just
performing the exchange $x_{i(j)}\leftrightarrow y_{i(j)}$ and
reversing the sign of the phases.

By inspecting (\ref{28}), we immediately see that the $|\vec x -
\vec y|$-factors completely cancel out and we conclude that the
$G^{+(-)}$ are dimensionless as they should. Counting the
$\mu_0$-factors in the above expression we also see that there is an
overall $\mu_0^{-2}$. Thus, inserting this result in (\ref{26}), we
conclude that the regulating mass (IR regulator) $\mu_0$ completely
disappears from the off-diagonal components of the fermion
correlator. As we shall see below, however, this situation is
modified when we consider the asymptotic behaviour of these
functions.

We may now analyze the asymptotic large distance behaviour of the
off-diagonal components of the fermion correlator, (\ref{26}). As we
saw in the case of (\ref{21}), the phases in (\ref{28}) will vanish
in this limit. Taking this fact into account and shifting the
integration variables as
\begin{equation}
\vec x_i \rightarrow \vec x_i - \vec x , \ \ \ \ \ \ \ \ \ \ \ \
\vec y_i \rightarrow \vec y_i - \vec x
\label{32}
\end{equation}
we can see from (\ref{28}) that
\begin{eqnarray}
\fl G^+ \left(|\vec x-\vec y|\right) \stackrel{|\vec x-\vec y|
\rightarrow \infty}{\sim} - \mathcal{Z}^{-1}\
\sum_{n=0}^{\infty}\frac{\alpha^{(2n+1)}}{n!(n+1)!}\int
\prod_{i=1}^{n}(\rmd^2x_i\, f(\vec x_i)) \prod_{i=1}^{n+1}(\rmd^2y_i\, f(\vec y_i)) \nonumber \\
\times \frac{\prod_{i<j}^{n} [x_i , x_j] \prod_{i<j}^{n+1}  [y_i ,
y_j]}{ \prod_{i}^{n}\prod_{j}^{n+1}[ x_i , y_j]} \nonumber
\\ \times \left(\frac{\prod_{i}^{n} [x_i , 0]}{\prod_{i}^{n+1}[y_i , 0]}\right)^{\frac{1}{2}} \left(\frac{\prod_{i}^{n}
[x_i , y-x]}{\prod_{i}^{n+1} [y_i , y-x]}\right)^{\frac{1}{2}} .
\label{33}
\end{eqnarray}

In the large distance regime, again, we must rewrite the last factor
in the above expression in a form analogous to (\ref{21b}), in terms
of the fully regulated Green's function, namely
\begin{eqnarray}
\fl \exp\left\{-(+)2\left[\sum_{i=1}^{n}{\rm {K}}_0
\left[\mu_0\left(|\vec x_i-(\vec y - \vec
x)|^2+|\varepsilon|^2\right)^{\frac{1}{2}}\right] \right .\right
.\nonumber \\ \left .\left . -\sum_{i=1}^{n+1} {\rm {K}}_0
\left[\mu_0\left(|\vec y_i-(\vec y-\vec
x)|^2+|\varepsilon|^2\right)^{\frac{1}{2}}\right]\right]\right\}.
\label{33a}
\end{eqnarray}
Using, as before, the fact that ${\rm {K}}_0(x)\stackrel{x
\rightarrow \infty}{\longrightarrow} 0$, we conclude that the above
expression tends to one for $|\vec y-\vec x| \rightarrow \infty$.
Counting the $\mu_0$-factors in the remaining terms of (\ref{33}),
taking into account (\ref{7}) and (\ref{20}), we find that they
completely cancel out. Therefore, inserting this result in
(\ref{26}) and renormalizing the fields as in (\ref{24}), we get
\begin{eqnarray}
\fl \langle\psi_{1(2)}(\vec x)\psi_{2(1)}^\dagger(\vec
y)\rangle_{\rm{R}} \stackrel{|\vec x-\vec y| \rightarrow
\infty}{\sim} \lim_{\varepsilon\rightarrow 0}\lim_{f(z)\rightarrow
1}\lim_{\mu_0\rightarrow 0} -(+)\rmi \ \mu_0^2|\vec x-\vec
y|^{\frac{3}{4}} \kappa(\varepsilon)^{+(-)} , \label{38}
\end{eqnarray}
where $\kappa(\varepsilon)^+$ is given by (\ref{33}), after removing
the last factor (notice that this is independent of $\mu_0$).

Using the Coleman's prescription that the mass (IR) regulator should
be eliminated first \cite{col}, we finally get
\begin{eqnarray}
\langle\psi_{1(2)}(\vec x)\psi_{2(1)}^\dagger(\vec
y)\rangle_{\rm{R}} \stackrel{|\vec x-\vec y| \rightarrow
\infty}{\longrightarrow} 0 , \label{39}
\end{eqnarray}
which coincides with the result for the off-diagonal components of
the fermion correlator in the massless Thirring model. These
components of the correlator vanish, in that case, because of
chirality conservation that exists in a massless fermionic theory.

We can understand physically, in terms of the CG picture, the reason
why the off-diagonal correlators vanish in the large distance
regime. The neutrality of the gas is responsible for the complete
cancelation of the IR regulator $\mu_0$. In the case of the diagonal
components, two external charges of {\it opposite} sign are
introduced for the description of the correlation function. In the
large distance regime, the external charges are removed to infinity
and decouple from the gas, since the fully regulated 2D Coulomb
interaction vanishes at large distances. Removing these charges to
infinity leaves a gas that remains neutral. The $\mu_0$-factors are
totally canceled out and the correlators are finite, as we can see
from (\ref{25}). Conversely, for the off-diagonal components, two
external charges of {\it the same} sign are introduced in the
system. These, together with the gas charges, form a neutral system.
When we remove the external charges to infinity, for describing the
large distance regime of the correlator, a non-neutral gas is left
after the decoupling of these charges. Then, the IR regulator
$\mu_0$ no longer cancels out and forces the correlation functions
to vanish for $\mu_0 \rightarrow 0$.

These results clearly expose the fact that the mass term of the MTM
becomes irrelevant at $\beta^2 = 8 \pi$.

\section{Concluding Remarks}

We would like to comment on the prescription adopted concerning the
regulators. When studying the asymptotic behaviour of the
correlation functions, we always take the limit $|\vec x-\vec y|
\rightarrow \infty $ firstly. Then, following \cite{col}, we take
the regulators out in the order: 1) $\mu_0 \rightarrow 0$; 2)
$f(\vec z) \rightarrow 1$ and 3) $\varepsilon \rightarrow 0$. This
leads, as we have seen, to the correct asymptotic limit of the
massive fermion correlators. For finite $|\vec x-\vec y|$,
conversely, we have shown that the $\mu_0$ regulator completely
cancels out and the limit $\mu_0 \rightarrow 0$ can be taken safely.
Nevertheless, when removing the UV regulator $\varepsilon$, we must
be careful because of the singularities that will appear due to the
short-distance Coulomb interaction of point charges. This has been
studied in detail for $4\pi \leq\beta^2 < 8\pi$  and it was shown
that the singularities that appear at multipole thresholds may be
absorbed by a subtractive renormalization of the ground-state energy
\cite{lima}. Nevertheless, the conjectured existence of a sequence
of phase transitions coinciding with these multipole thresholds in
the region $4\pi \leq\beta^2 < 8\pi$ \cite{ben,gall}, has been later
on denied \cite{mf,sr5}. For $\beta^2 \geq 8\pi$, however, the UV
problem becomes extremely complicated and, as far as we know,
remains unsolved. Consequently, only the large distance regime
($|\vec x-\vec y| \rightarrow \infty)$ of the MTM, which has been
studied here, can be considered sensible in this region of the
coupling $\beta$.

\ack

This work has been supported in part by CNPq and FAPERJ. LM was
supported by CNPq and ECM was partially supported by CNPq.


\section*{References}
\begin{thebibliography}{23}
%
\bibitem{MM1} Mondaini L and Marino E C 2005 \emph{J. Stat. Phys.} \textbf{118} 767
%
\bibitem{KT} Kosterlitz J M 1974 \emph{\JPC} \textbf{7} 1046
%
\bibitem{tsvelik} Tsvelik A M 1995 \emph{Quantum Field Theory in Condensed Matter Physics}
(Cambridge, UK: Cambridge University Press)
%
\bibitem{sgcg} Samuel S 1978 \emph{\PR}D \textbf{18} 1916
%
\bibitem{col} Coleman S 1975 \emph{\PR}D \textbf{11} 2088
%
\bibitem{mand} Mandelstam S 1975 \emph{\PR}D \textbf{11} 3026
%
\bibitem{zz} Zamolodchikov A B and Zamolodchikov Al B 1979 \emph{\APNY}
\textbf{120} 253 \\ Korepin V E 1980 \emph{Commun. Math. Phys.}
\textbf{76} 165
%
\bibitem{faber} Faber M and Ivanov A N 2001 \emph{Eur. Phys. J.} C \textbf{20} 723
%
\bibitem{faber2} Faber M and Ivanov A N 2003 \emph{\JPA} \textbf{36} 7839
%
\bibitem{juricic} Juricic V and Sazdovic B 2004 \emph{Eur. Phys. J.} C \textbf{32} 443
%
\bibitem{dv} Destri C and de Vega H J 1995 \emph{\NP}B \textbf{438} 413
%
\bibitem{sr1} Lukyanov S and Zamolodchikov A 1997 \emph{\NP}B \textbf{493} 571
%
\bibitem{sr2} Lukyanov S and Zamolodchikov A 2001 \emph{\NP}B \textbf{607} 437
%
\bibitem{babu1} Babujian H, Fring A, Karowski M and Zapletal A 1999 \emph{\NP}B \textbf{538} 535
%
\bibitem{babu2} Babujian H and Karowski M 2002 \emph{\NP}B \textbf{620} 407
%
\bibitem{babu3} Babujian H and Karowski M 2002 \emph{\JPA} \textbf{35} 9081
%
\bibitem{sr6} Samaj L and Jancovici B 2002 \emph{J. Stat. Phys.} \textbf{106} 323
%
\bibitem{sr7} Samaj L 2003 \emph{\JPA} \textbf{36} 5913
%
\bibitem{sr4} Samaj L and Travenec I 2000 \emph{J. Stat. Phys.} \textbf{101} 713
%
\bibitem{sr5} Kalinay P and Samaj L 2002 \emph{J. Stat. Phys.} \textbf{106} 857
%
\bibitem{sr3} Alastuey A and Cornu F 1992 \emph{J. Stat. Phys.} \textbf{66} 165
%
\bibitem{amit} Amit D J, Goldschmidt Y Y and Grinstein G 1980 \emph{\JPA} \textbf{13} 585
%
\bibitem{zj} Zinn-Justin J 2002 \emph{Quantum Field Theory and Critical Phenomena}
(New York: Oxford University Press Inc.)
%
\bibitem{KosT} Kosterlitz J M and Thouless D J 1973 \emph{\JPC} \textbf{6} 1181
%
\bibitem{jose} Jos\'{e} J V, Kadanoff L P, Kirkpatrick S and Nelson D R 1977 \emph{\PR}B \textbf{16} 1217
%
\bibitem{giam} Giamarchi T and Schulz H J 1989 \emph{\PR}B \textbf{39} 4620
%
\bibitem{le} Luther A and Emery V J 1974 \emph{\PRL} \textbf{33} 589
%
\bibitem{klaiber} Klaiber B 1968 \emph{Lectures in Theoretical Physics} vol~10A ed A O Barut and W E Brittin (New York: Gordon and
Breach)
%
\bibitem{grad} Gradshteyn I S and Ryzhik I M 2000 \emph{Table of Integrals, Series, and
Products} ed A Jeffrey and D Zwillinger (San Diego: Academic Press)
%
\bibitem{ms} Marino E C and Swieca J A 1980 \emph{\NP}B \textbf{170[FS1]} 175
%
\bibitem{arf} Arfken G 1970 \emph{Mathematical Methods for Physicists} (New York: Academic Press, Inc.)
%
\bibitem{lima} Lima-Santos A and Marino E C 1989 \emph{J. Stat. Phys.} \textbf{55}
157
%
\bibitem{ben} Benfatto G, Gallavotti G and Nicol\`{o} F 1982 \emph{Commun. Math. Phys.} \textbf{83}
387 \\ Nicol\`{o} F 1983 \emph{Commun. Math. Phys.} \textbf{88} 581
\\ Nicol\`{o} F, Renn J and Steinmann A 1986 \emph{Commun. Math. Phys.} \textbf{105}
291
%
\bibitem{gall} Gallavotti G 1985 \emph{\RMP}
\textbf{57} 471
%
\bibitem{mf} Fisher M E, Li X and Levin Y 1995 \emph{J. Stat. Phys.}
\textbf{79} 1
%
\end{thebibliography}
\end{document}